# Evaluation of Induced Transmembrane potential on Membrane Poration through Molecular Dynamics Simulation and Analytical Calculation

**Shadeeb Hossain[1][ORCID: 0000-0002-5224-7684]**

[1]School of Engineering and Technology, Central Michigan University, MI, USA.

**\*Corresponding Author: shadeebhossain@yahoo.com**

**Evaluation of Induced Transmembrane potential on Membrane Poration through**

**Molecular Dynamics Simulation and Analytical Calculation**

**Abstract**

A molecular dynamics (MD) simulation is used to quantitatively analyze the induced membrane potential for an applied external field varied between 0.4 V/nm to 2.0 V/nm. The change in the electrostatic potential in the DPPC is directly correlated to the membrane permeability. The effect of the decrease in external conductivity on the DPPC is also evaluated and the  analytical results are compared with the simulation. The correlation between the electrostatic potential of the DPPC and the total dipole are compared, and a positive correlation is identified until saturation. This is because the membrane permeability factor plays a dominant role to control reversible and irreversible electroporation. The obtained dipole parameters through simulation for various electric field allows for an accurate determination of the quantitative changes in the membrane and external conductivity during the process of electroporation.

**I. Introduction**

Electroporation is a physical process of structural rearrangement of the membrane due to application of high magnitude electric pulses [1-3]. Nanometer sized pores are formed on the membrane surface accounting for increased conductivity of the cell [4-6]. Electroporation has miscellaneous applications including cancer therapy drug delivery, gene transfer and therapeutic destruction of cancerous cell or other microorganisms [7-11]. There are advantages of electroportaion over other therapeutic treatment including: (i) flexibility of pulse duration that can allow manipulation of membrane permeability (ii) negligible Joule heating (iii) electrochemotherapy (ECT): a clinical practice to assist the delivery of chemotherapeutic drug at tumor site [12-14]. Electric pulses in nanosecond range with magnitude as high as 10 kV/cm (or 0.001 V/nm) can allow active cell manipulation [15, 16]. Recent studies have shown that the application of these electric pulses affect the integrity of intracellular structures and triggers cytochrome c release (it is a protein present in mitochondria). This increases intracellular calcium level that facilitate gene transfer or drug uptake [17, 18].

Reversible electroporation has the advantage of transient pore formation with active resealing occurring in milli-second (to upto a few seconds) range thereby preventing osmotic imbalance [19]. This increases its proclivity for drug delivery and gene transfer. Electrochemotherapy is an efficient, *once-only* treatment required (with a response rate of 72% to 100%) for treating cancer patients [13,20]. The electric pulses can cause induced change in transmembrane potential; thereby increasing the membrane permeability through nanopores that can allow molecules with molecular weight of approximately 30 kDa to diffuse into the cell [20,21]. Sera et al. (2008) performed a study to analyze that electrochemotherapy has vascular disrupting action [22]. The electric pulses were of the following parameters: 1.04 kV, 0.1 ns with 8 pulses operating at 100Hz frequency. The electro chemotherapeutic drug was bleomycin (BLM) to treat subcutaneous sarcoma tumor in a mice specimen. Almost 38% tumor was cured reflecting good anti-tumor effect within 22 days. However, clinical trial with electrochemotherapy was performed by Belehradek et al. (1993) [23]. The electroporation parameters were 1.3 kV/cm (or 0.00013 V/nm), 0.1 ns with 4 to 8 pulses. The therapeutic drug was hydrophylic bleomycin that was used to treat 8 patients with 40 permeation nodules of head and neck squamous cell carcinomas [23]. A positive anti-tumor response was found with 57% of the nodules that responded within a few days.

Irreversible electroporation is a monotherapy used for cell ablation and has therefore potential application in malignant cell treatment [13,24-26]. There is a threshold range for electric field application that ensures membrane resealing or irreversible permeabilization eventually leading to cell apoptosis [27]. Miller et al. (2005) performed an in vitro study of irreversible electroporation on 600 μL of cell suspension of human hepatocarcinoma cell (HepG2) with a concentration of one million cells per mm on a 2 mm gap cuvette [28]. The corresponding electric field parameters for cancer cell ablation was: 1.5 kV/cm (0.00015 V/nm), 300 μs with three sets of 10 pulses. In a similar study performed by Rubinsky et al. (2007), the concept of irreversible electroporation was studied on 35 lesion tissues produced from the liver of 14 pigs [29]. The electric field parameters were: 0.6 kV/cm ($6 \times 10^{-5}$ V/nm) for 0.1 ns with 8 pulses separated on an interval of 100 μs. At the end of the study period of 14 days the ablated lesion area was analyzed to be replaced by fibrous scar tissue.

The discrepancy of electric field parameters depending on the cell nature requires an in-depth understanding of the process. Further in vivo studies require better understanding through both molecular dynamics simulation and numerical analysis on the dependence of influential parameters governing electroporation. Understanding the theoretical and computational model of reversible electroporation can allow comprehensive understanding on determining the threshold for therapeutic treatments like the synergistic effect of shock impulse and electric pulse application [30-32].

Molecular Dynamics (MD) simulations can be an effective methodology to analyze the changes in the cell membrane permeability under the influence of an external electric field. They allow modeling the trajectories of atoms by integrating Newton's equation of motion for any defined system. This generated simulation has the advantage of providing the time dependent response to the electric field perturbation technique. This modeling method will provide the necessary mathematical information regarding the conformational atomic and molecular arrangement when influenced by the electric field. The advantages of opting for MD simulation include: (i) detailed atomic/molecular level information (ii) ability to control a single influencing factor and thereby analyze its interatomic conformation (iii) alternative to study rapid perturbation techniques such as electric pulses in pico-second range and obtain quantitative data at atomic level resolution.

Tarek (2005) performed MD simulation of lipid bilayer to study the effect of transverse electric field [33]. The system required 0.5 V/nm and 1.0 V/nm electric field magnitude to induce electroporation by formation of water channels across the membrane. The system was also studied for DNA strand migration into the interior of the membrane due to electric field application. Tieleman (2004) et al. investigated the mechanism of pore formation through MD simulation of varied strengths [34]. The bilayer of 26x29 nm produced several pores of diameter 10 nm for an applied electric field of 0.5 V/nm. However, for a field strength of 0.4 V/nm, (corresponding transmembrane potential of 3V), the system only produced a single pore. Casciola et al. (2014)

investigated the effect of ms and ns pulse duration on the required electroporation threshold [35]. It was realized that for ns pulses the pores were hydrophilic and for ms pulses it was hydrophobic. Similarly, the electroporation threshold increased linearly with cholesterol content for ns pulses.

The focus of our model is to investigate the electroporation threshold using a Dipalmitoyl phosphatidylcholine (DPPC) bilayer to replicate a cell membrane model. The changes in the hydrophobic nature of the DPPC is analyzed to determine the electroporation threshold, $U_{threshold}$. Similarly, the pore permeabilization is analyzed to determine its reversible nature. It is also compared with parameters particularly, the electrostatic potential variation along the DPPC region and information is obtained about the changes in the total dipole of the model along the transverse applied field direction.

For simulating our model, leapfrog algorithm was used to integrate the equation of motion. The simulation time was set to 2 fs and the total simulation period was 100 ps to analyze the intermolecular/interatomic conformation. DPPC has been used in previous literature [30-32] to model the influence of perturbation on a biological cell. It has a hydrophilic head and hydrophobic tail. This amphipathic lipid can therefore replicate the semi permeable nature of cell membrane. The primary objective of the MD simulation is to analyze the effect of varied intensity electric field on the hydrophobic arrangement of DPPC layer. Fig.1 shows the simplified schematic of the simulation model for this study. Electric field of varied intensity is applied across the z-direction of the simulation model and the changes in the hydrophobic nature of the DPPC layer is studied. Also, relevant information regarding the electrostatic potential across the hydrophobic region of DPPC and the variation in dipole charges are all studied to obtain atomic level understanding of the process. The obtained data is then compared with the mathematical model of electroporation to comprehend the influencing factors at both atomic/ molecular level and as a spherical single cell model.

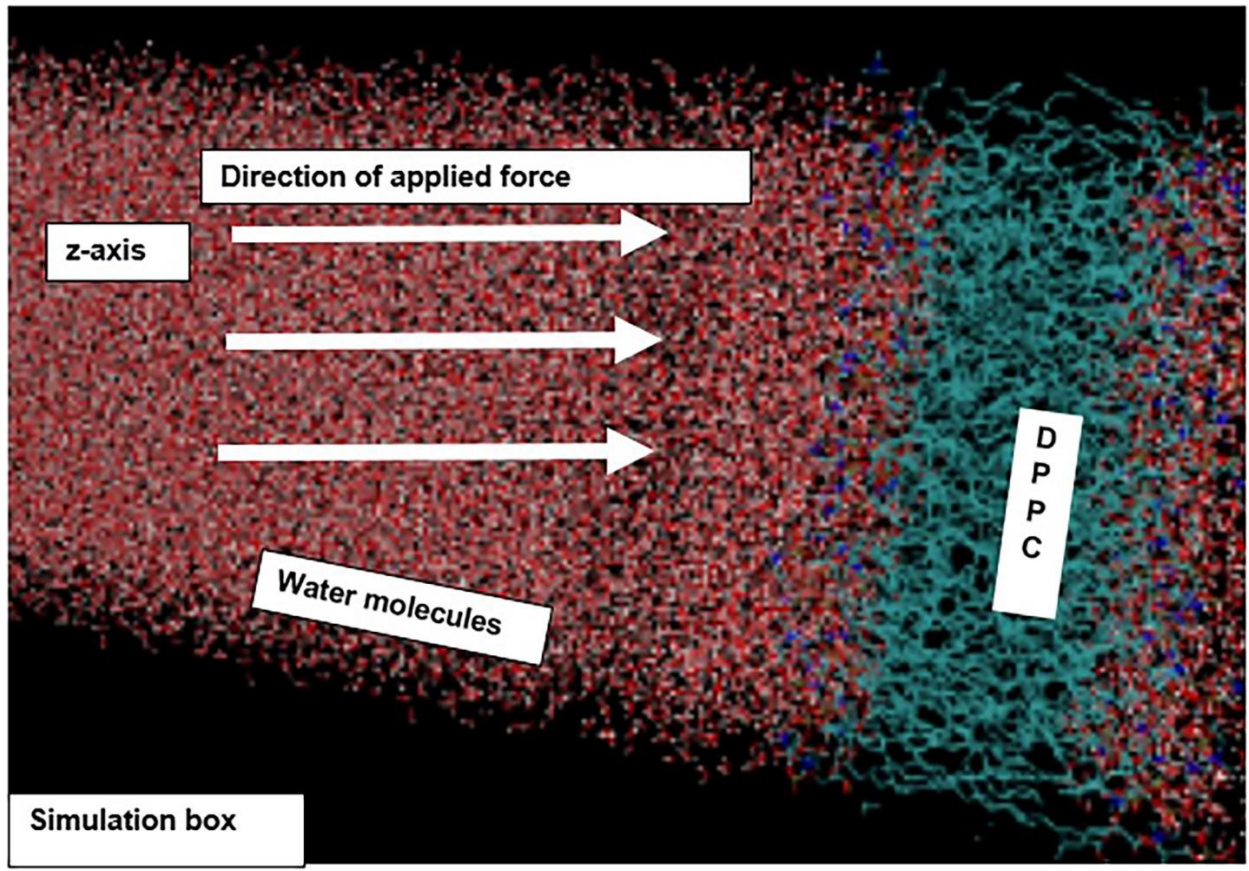

**Fig.1: Schematic of the MD simulation to analyze the application of electric field on DPPC layer. The magnitude of electric field was varied from 0 to 2.0V/nm for a simulation period of 100ps.**

Joshi et al. (2004) modelled the potential across each node of a double shelled spherical cell model based on the current continuity equation [12]. The quantitative time dependent model analyzed the transmembrane potential change across the inner and outer membranes across various electric field intensities and pulse duration in ns range. The model allowed to discern malignant and healthy cells. For instance, their model predicted that 45-150 kV/cm (0.0045 V/nm -0.015 V/nm) electric field intensity in 10-50 ns pulse could annihilate malignant tonsillar B- cells whereas it does not damage normal B-cells. Delemotte et al. (2008) modelled ion transport through biological transmembrane channels [36]. They studied the transmembrane potential generated by asymmetric distribution of ions across the membrane. The computational model for high transmembrane potential of 500 mV-3V induced membrane electroporation. The effect of charge imbalance and its intrinsic property of self-readjustment to avoid disruption was investigated. Hibino et al. (1991) analyzed the transmembrane potential induced in sea urchin egg on application of microsecond electric pulses across the cell [37]. The time dependence spatial resolution of the induced transmembrane potential aligned in accord with the theoretical model. The group analyzed membrane conductance due to electroporation of the cell. Freeman et al. (1994) analyzed transient aqueous pore theory to determine the fraction of membrane area occupied by pores [38]. The results showed reversible electrical breakdown of bilayer membrane on having the transmembrane potential reaching 1V which also supported a rapid escalation of pore filled membrane area. For moderately higher transmembrane potential irreversible electrical breakdown can occur causing the pore to expand to membrane boundary.

This paper hence analyzes the change in membrane permeability due to application of external field across a range of various intensities. Literature models to analyze quantitatively the change in induced membrane potential is derived and compared with computational modeling. The MD simulation compliments the understanding of various physical parameters that support the hypothesis.

The focus of this paper is to determine the correlation between applied electric field and the corresponding property changes in (ii) *electrostatic potential* (ii) *dipole* and (iii) *pressure changes* that occurs along the DPPC layer through computational modeling. The concept of electroporation in cell membrane due to external applied field is correlated to transient change in membrane potential but a critical correlation between measurement of the instantaneous potential change along the hydrophobic DPPC and associated dipole changes in the molecules during the external stimuli is imperative to understanding the biophysical property changes in DPPC. The relative quantitative measure will allow to identify the threshold of external field to induce DPPC permeability as opposed to its hydrophobic region. Proposed models by *(i) Paul and Schwan (1959) (ii) Maxwell (1873) and (iii) Takashima (1989)* provides an understanding on the transient changes in conductivity and membrane potential on a macroscopic level but MD simulations allows scope to understand the associated forces and property changes in molecules at microscopic level.

## II. **Theoretical Background**

Understanding the process or the parametric factors that influence the process of electroporation will allow scope for *in vivo* applications [39-42].

### A. **Modelling the Cell Membrane**

The cell membrane has a double layer of lipids and proteins [43,44]. The lipid bilayer consists of phospholipid molecule with a triglycine hydrophilic head attached with two hydrocarbon hydrophobic tail, making it amphiphilic [44]. To understand the cell membrane, it can be compared to an equivalent combination of resistance and capacitance. The capacitance arises due to the dielectric nature of the cell with the accumulation of charged particles across the bilayer lipid surface. However, the resistance accounts for the conductivity of ion channels. The biological cell consists of several ions including $Na^+$, $K^+$, $Cl^-$, $Ca^{2+}$, etc. [45] and it is the ion concentration gradient that contributes to membrane potentials [46]. The intracellular $K^+$ ion concentration is higher whereas the extracellular $Na^+$ and $Cl^-$ concentration is relatively more. The membrane potential is therefore calculated $\Delta U = U_{external} - U_{internal}$ , where $U_{external}$ and $U_{internal}$ are the external and internal membrane potential respectively. Fig. 2 shows a simple schematic of a spherical biological cell and its electrical equivalent circuit diagram. Time constant is a fundamental factor for evaluation of the time scale during electroporation. It is represented as $\tau = RC$, where R and C are membrane resistance and capacitance respectively. Time constant helps to predict the implications for in vivo studies with cell-electric pulse interaction. It allows to quantitatively analyze the impact of cellular damage during microsecond E-pulse application [47].

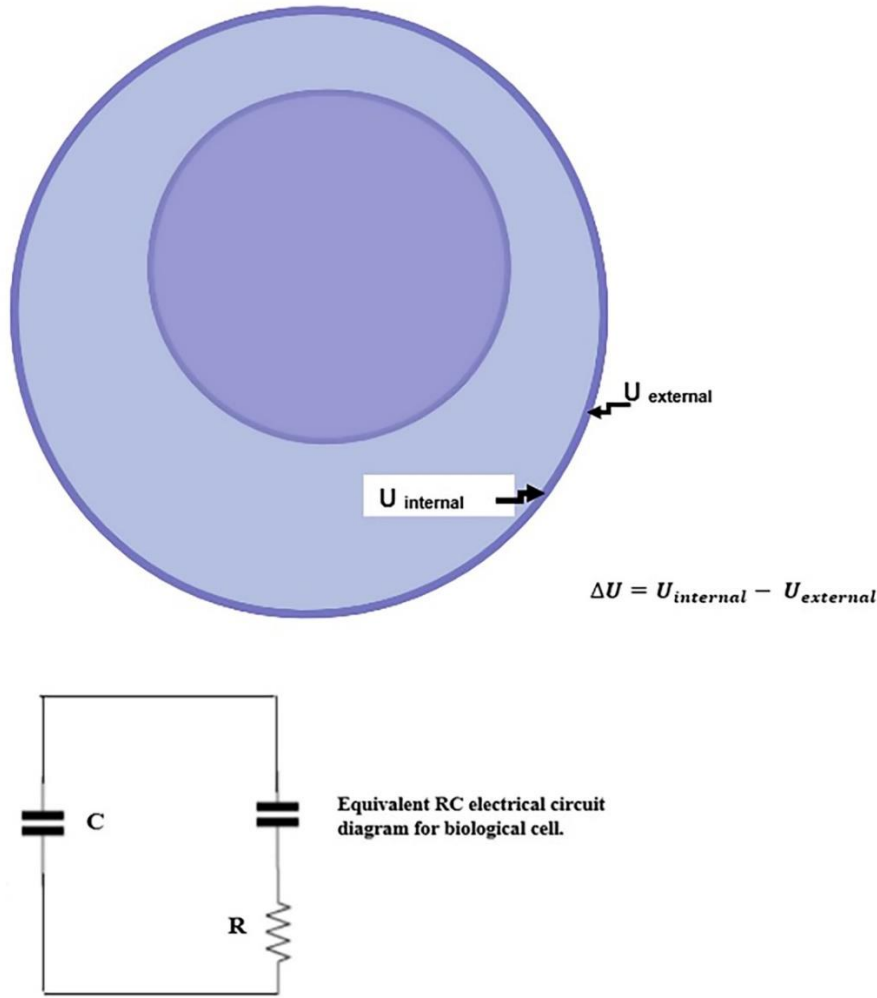

Fig. 2: Schematic of a spherical biological cell and its electrical RC equivalent circuit diagram.

## B. Electrophoresis for charge segregation on plasma membrane

The application of an external field facilitates lateral charge mobility and redistribution across the plasma membrane. This process is called perimembrane electrophoresis [48]. Jaffe et al. (1977) characterized the polarizability of particles across the membrane using equation (1):

$$Q = \left(\frac{m}{D}\right)\left(\frac{V}{6}\right) \tag{1}$$

where $Q$ is the degree of polarization, $m$ is the electrophoretic mobility, $V$ is the steady state voltage drop across the cell and $D$ is the diffusion constant.

Hence for a polarizability of 0.1, the corresponding steady state voltage drop is 0.8 mV. For a higher polarizability, if the distribution of particles across the membrane surface is increased to half proportion polarized, the magnitude of steady state voltage drops increases of 4 mV. Hence if the surface particles were all concentrated at one pole of the membrane, V increases to 7.8 mV. However, the duration of this equilibrium theory is applicable for prolonged field duration of approximately 2.7 hours. The steady state transmembrane potential of a cell is between -40 to -70 mV and hence the corresponding ratio of diffusion constant and electrophoretic mobility is expected at $23.3 \times 10^{-3}$ with the assumption that the molecules in the plasma membrane are half polarized. During the process of electroporation, which involves the application of external electric field (indifferent from the bioelectric fields of the charged components in the cell); there is a transient increase in the membrane permeability.

## C. Stochastic Electroporation model

*Transmembrane potential*

The resting transmembrane potential (measured in tens of milli volts) on cell surface is due to the distribution of ion pumps and the gated channels of the system [49]. However, the external perturbation by an electric field influences the intrinsic electrical properties of the cell causing increased permeability and conductivity of the cell. This arises due to the superposition of intrinsic field of cell with the applied external field. Therefore, there is a transient escalation in the membrane potential and is called *induced transmembrane potential*. There is a proportionality between field strength and the transient increase in membrane potential, Δ ψ [50-52]. The critical range for reversible permeabilized pore is 250 mV to 1 V [53,54] but the magnitude is contingent on the cell type and other intrinsic properties [55].

Pauly and Schwan (1959) described the increase in induced membrane potential according to equation (2) [56]:

$$\Delta\psi_m = \frac{3}{2} ER \, Cos \, \theta \, ( \, 1 - e^{-t/\tau_m})  \tag{2}$$

where E is the external electric field, R is the radius of the cell (approximately 10µm), $\theta$ is the degree angle of the applied field measured from the center of the cell, t is transient time and $\tau_m$ is the charging constant of membrane (discussed briefly in Section-*Modelling the Cell Membrane*).The above equation has an upper threshold frequency of 1MHz and pulse duration of 1 µs [57]. The equation by Schwan (1983) later accounts the angular frequency, *w* of the external field [58].

$$\Delta\psi_m = \frac{\frac{3}{2} ER \, Cos \, \theta}{1+iwaC(\rho_e+\rho_i/2)}  \tag{3}$$

where C is the membrane capacitance/area (approximately $10^{-2}$ F/m$^2$), a is membrane resistance $\rho_i$ is the resistivity of the interior of the cell and $\rho_e$ is the resistivity of the electrolyte (approximately 1 ohm.m).

The application of high frequency pulses affects the intracellular structure of cell and actuate cytochrome c release, which facilitates intracellular calcium level [59,60]. The change in ion concentration could have contributed to change in membrane potential. There is a layer of surface membrane conductance which is characterized by the enhancement of ion density near the vicinity of cell [61]. This could have been due to several reasons including the one listed above [59,60]. Other possible explanations include the layer of glycocalyx, surrounding the cell (which has fixed charges) is more likely to attract ions and influence membrane potential. Grosse and Schwan (1992) derived an expression for membrane conductance, equation 4, that includes the component of surface conductance, $G_s$ [61]. Due to the increase in membrane conductance due to application of transient electric field, the membrane potential and time constant decreases [61]. It has been reported that the magnitude of $G_s$ is of the order of $10^{-8}$ S in literature [62] (though typical value is of the order of $10^{-9}$ S).

$$\Delta\psi_m = \frac{\frac{3}{2}ERCos\,\theta}{1+\frac{\rho_e G_s}{R}+iwRC(\rho_e/2+\rho_i+\frac{\rho_i\rho_e G_s}{R})} \tag{4}$$

*Membrane Conductance*

Analyzing the membrane conductance due to electroporation helps in comprehending the escalated transient permeabilization process. Hibino et al. (1991) analyzed the large electrical conductance due to induced transmembrane potential [37]. Their results showed a membrane conductance of 1 S/cm$^2$ within 2 µs from the application of external field. Pavlin et al. (2003) used analytical calculation using effective medium theory to calculate the effective conductivity of a cell, σ [63]. Maxwell (1873); Takashima (1989) obtained the following equation (5) involving effective conductivity [63-65].

$$\frac{\sigma_e - \sigma}{\sigma + 2\sigma_e} = f \frac{\sigma_e - \sigma_p}{2\sigma_e + \sigma_p} \tag{5}$$

where $\sigma_p$ is the conductivity of the particles, $f$ volume fraction of the particles in the medium and $\sigma_e$ is conductivity of external medium. Pauly and Schwan (1959); Dukhin (1971) obtained the following equation (6) for conductivity of particles [63,66,67].

$$\sigma_p = \sigma_m \frac{2(1-v)\sigma_m + (1+2v)\sigma_i}{(2+v)\sigma_m + (1-v)\sigma_i} \tag{6}$$

where $\sigma_m$ is the membrane conductivity, $\sigma_i$ intrinsic conductivity of cytoplasm and $v$ is calculated $v=(1-d/R)^3$, d=membrane thickness and R is the radius of the cell.

Hence, by comparing equation (5) and (6) the effective conductivity can be obtained. Table-1 compares the magnitude of the parameters to calculate effective conductivity under normal physiological conditions and when influenced by electric field.

**Table-1 compares the magnitude of the parameters to calculate effective conductivity under normal physiological conditions and when influenced by electric field.**

| Parameters | Magnitude | Physiological condition |
|---|---|---|
| $\sigma_e$ | 0.5 S/m [63] <br><br> 0.158 S/m[68,69] | Both normal condition |
| $\sigma_i$ | 0.5 S/m [63] | Both normal and under electric field influence |
| $\sigma_m$ | $2 \times 10^{-4}$ S/m [37],[63] | During electroporation |
| $\sigma_m$ | $10^{-5}$ S/m [63] | No field |
| R | 10 µm | Both normal and under electric field influence |
| d | 5 nm | Both normal and under electric field influence |

## II.   SIMULATION PARAMETERS

The contribution on the lipid bilayer permeability during interaction of applied electric field was simulated based on atomistic MD simulation. The simulated cell was of the following dimension: 6.902 x 7.405 x 9.504 nm. The DPPC membrane was built with 128 DPPC layer surrounded by 3655 water molecules. The time step was set to 2fs and the total simulation period was 100ps. The simulation was performed using GROMACS v.4.6.7 software package and was carried in NPT (isothermal-isobaric) ensemble at pressure=10 atm and temperature= 320K.  Temperature and pressure coupling were performed using Parrinello-Rahman coupling method. The MD simulation was performed using leapfrog algorithm to numerically update the position and velocities of the particles in the system.. A cut off distance for Lennard –Jones interaction was set to 1nm and particle-mesh Ewald method was used for the electrostatic interaction with a cut off distance of 1nm. Periodic boundary condition was applied along x, y and z direction. GROMACS was used for the quantitative study of the field induced membrane effect in association with VMD for the visual data of the study.

The effect of different magnitude of electric field on the membrane permeability can be determined with its application along the z direction. The electric field was applied perpendicular to the membrane as shown in the schematic in Fig. 1. The magnitude of the electric field was varied from 0V/nm to 2.0V/nm with an increment interval of 0.4V/nm during each simulation period. This method is analogous to the application of a short electric pulse on a cell membrane and the permeability change was studied through analyzing the osmotic effect of water in the hydrophobic DPPC layer. Fig. 3 shows the DPPC bilayer used for simulating the system along with the hydrophobic DPPC region along the z-axis.

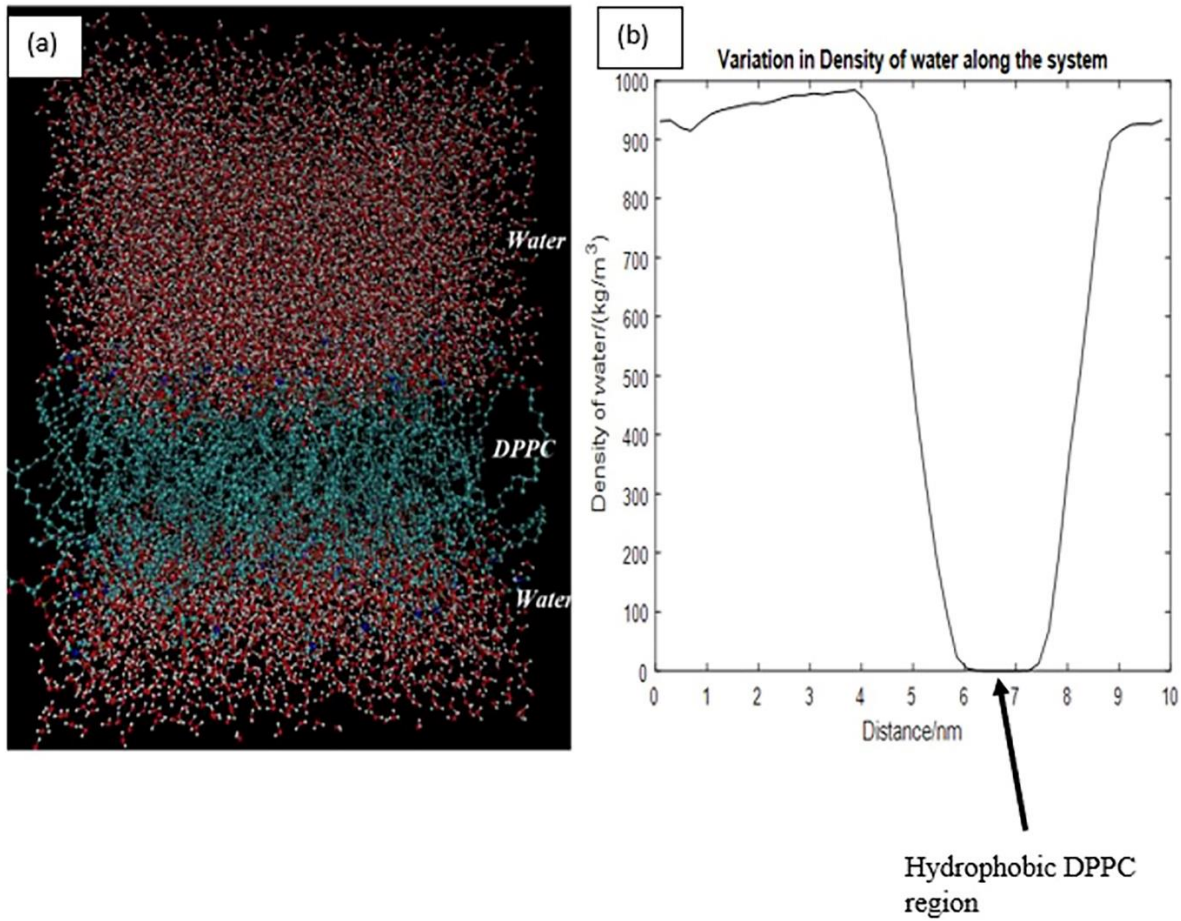

**Fig. 3: (a) DPPC bilayer VMD visual interface (b) variation in water density along the z-axis, showing the hydrophobic DPPC region.**

### III. Results and Discussion

Electric pulses of field intensities 0.2 V/nm and 1.2 V/nm with pulse duration of the order in microsecond was analyzed to determine the membrane potential change. It is assumed that the cell membrane has a resistive-capacitive electrical equivalent model and the electroporation in the cell leads to increased ionic conductivity and pore formation when the membrane potential exceeds a critical threshold magnitude. Short microsecond pulses have the advantage over millisecond pulses because of the fast-rising pulses would create a 'window' period showing a large non-equilibrium transient membrane potential change  [12]. The membrane time constant, $\tau$ is experimentally determined to be 1 µs (Kinosita et al. 1988) and hence the steady state membrane potential is acquired at that time instant [70]. The time dependence of the change in membrane potential is calculated from equation (2) and shown in Fig. 4. The graph shows a direct dependence of field intensity to transient increase in membrane potential and reaches a stable membrane potential at $\tau$. This is an important parameter as the change in membrane potential is more likely to be influenced by cell nature and its electrical parameter $\tau$. The angular dependence of the field is kept at zero to calculate the maximum change in the approximate membrane potential. The graph in Fig. 4 is generated from equation (2) is applicable for cell sizes in µm for the frequency of kHz to MHz range. The elevated membrane potential under higher field of 1.2 V/nm indicates a greater entry of $Ca^{2+}$ . When the membrane reaches a critical threshold voltage, it leads to pore formation and allows ionic current conduction into the cell [37].

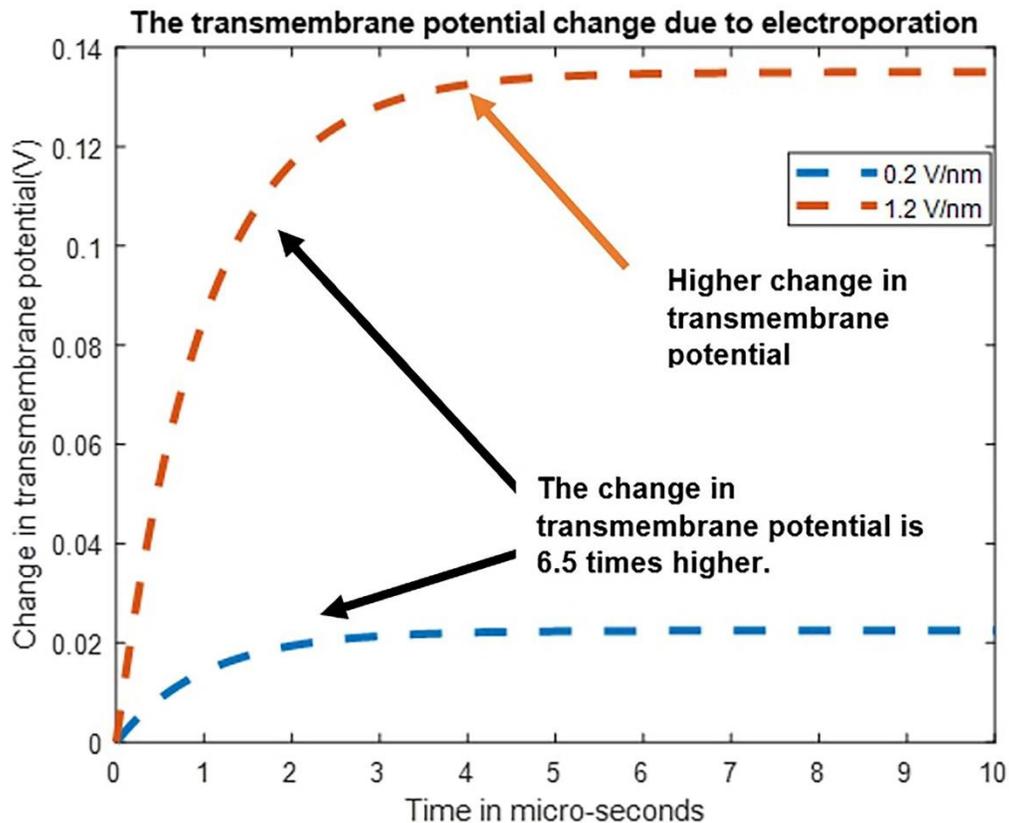

**Fig. 4 : Time dependence of the change in membrane potential at 0.2 V/nm and 1.2 V/nm for 1 μs pulse duration.**

The critical threshold membrane potential for the ionic current conduction is approximately 200 mv- 1 V ( [71-73]. Fig. 5 shows the potential change across a range of frequency from 1 kHz to 1 MHz. The trend is like that obtained by Gross and Schwan (1992) and the magnitude of the membrane potential is like obtained from Fig. 4 for field intensity of 1.2 V/nm [61]. Fig.5 shows a decrease in membrane potential at higher frequencies due to surface or membrane conductance (depending on cell size). The shielding effect by a highly conducting external wall or glycocalyx causes the membrane potential to decrease [61]. The skin effect at higher frequencies (upper MHz frequency range) causes the membrane resistance to increase; this could have accounted at molecular level through shielding effect of glycocalyx eventually leading to decreased membrane potential.

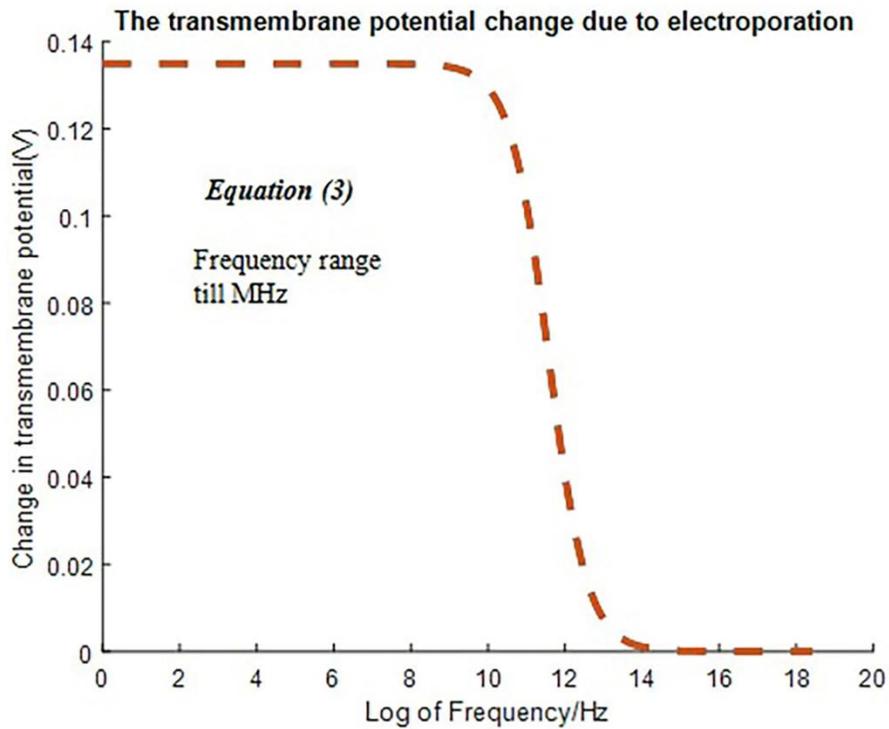

**Fig. 5 : Influence of membrane potential across different frequency range.**

A review article illustrates on the correlation of external conductivity, $\sigma_e$ on the electroporation phenomenon [74]. It summarized that in most studies, the electroporation efficiency was directly correlated to external conductivity; that is electroporation efficiency decreased as external conductivity decreases. However, they also reinforced that this sensitivity varied between experiments and on the assessment, technique utilized to measure permeability. In our study, we tried to measure membrane permeability via (i) change in transmembrane potential (ii) this eventually contributed to increased water concentration in hydrophobic region of DPPC (as discussed later in the simulation section). Fig. 6 compares the electroporation efficiency via measuring reactive component of the transmembrane potential using equation (3) and different external conductivity [63,68,69]. The nature of the graph agrees with the general hypothesis in [68,69,74] that is as the external conductivity decreased the electroporation efficiency decreased (measured via decrease in reactive component of the transmembrane potential). The transmembrane potential change was analytically assessed for field magnitude of 3000 V/cm (0.0003 V/nm) measured at an angular degree of $0^0$. If the applied electric field is above threshold magnitude, electroporation occurs contributing to an abrupt increase I membrane conductivity. The membrane conductivity depends on parameter, K which is calculated from external and internal conductivity [74]. Hence the induced membrane potential drop is significantly dependent on external conductivity. The graph in Fig. 6 in other words support that if the external medium conductivity decreases, a higher magnitude electric pulse is required to cause the same permeability. Hence, the external membrane conductivity of the cell is an influential factor in determination of the field strength required for electroporation.

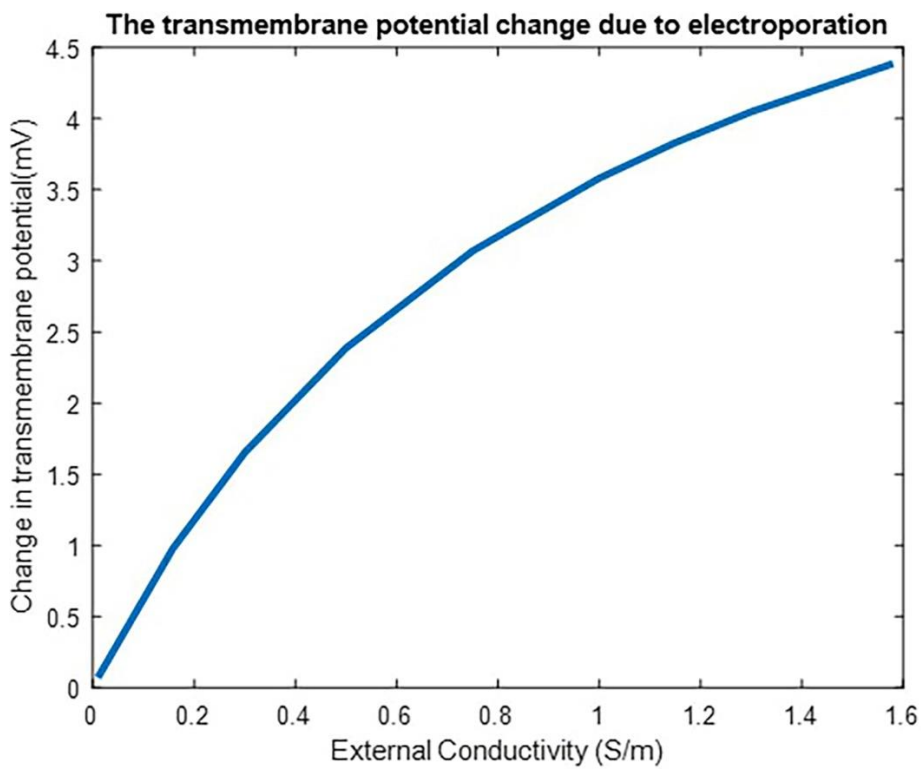

**Fig. 6: Analysis of the electroporation efficiency as a function of external conductivity (via measuring reactive component of the transmembrane potential).**

*Analysis of electrical potential across DPPC during MD simulation*

Fig. 7 compares the snapshot for the simulation model for electric field amplitude 1.2 V/nm. This time dependent visual data is obtained using Visual Molecular Dynamics (VMD) graphics software. At the beginning of the energy minimized simulation, it is evident that the water concentration (red/white water molecules) in the hydrophobic region of DPPC (green) is negligible. The primary focus of this simulation is to obtain the critical field value required to permeabilize DPPC and try to correlate with two parameters (i) change in electrostatic potential

(ii) dipole of the system. Pavlin and Miklavcic (2003) emphasized the significance of dipole term in their analytical calculation of effective conductivity using equivalent principle [63].

From Fig. 7 and 8 it is evident that there is a critical threshold electric field required for water channel to permeate through the hydrophobic DPPC. The nature of the cell is also a crucial parameter for determining this magnitude. As it is evident from the data in Fig. 7 and 8 that an approximate 1.2 V/nm is the required field intensity for induction of permeabilization. However, this critical threshold depends on the cell size. Electro permeabilization process is only triggered by characteristic threshold, $E_s$ =1.2 V/nm (for this model). This field intensity causes the DPPC simulated structure to attain the transmembrane potential difference allowing the membrane to become permeabilize. The field distribution on the stimulated structure is a function of *Cos θ*.

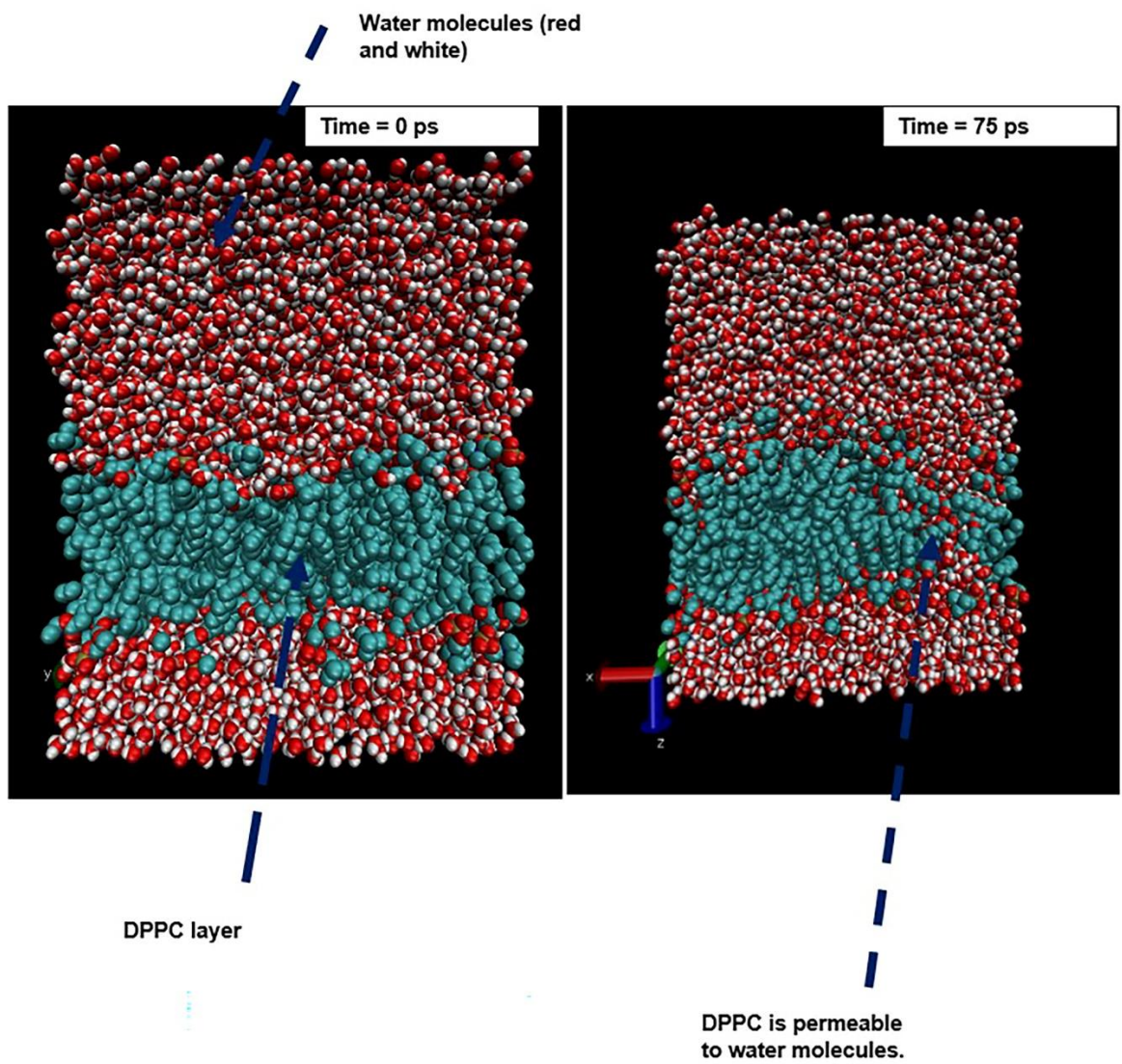

**Water molecules (red and white)**

Time = 0 ps

Time = 75 ps

**DPPC layer**

**DPPC is permeable to water molecules.**

**Fig. 7 : Time dependent snapshot of the simulation model for applied electric field of 1.2 V/nm.**

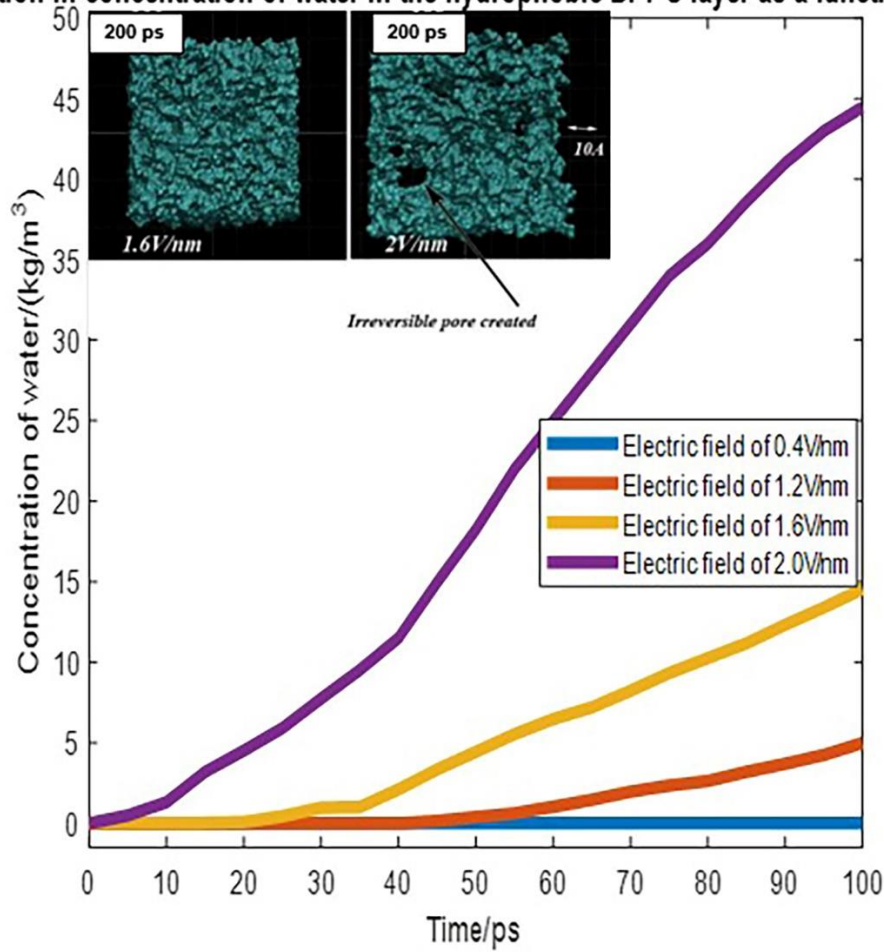

**Fig. 8: The variation in the density of water molecules along hydrophobic region of DPPC layer.**

Upon reaching the critical value of induced transmembrane potential, further escalation of electro-induced potential difference is limited by a decrease in its membrane permeability factor. This is realized through a saturation curve obtained in Fig. 9 for the electrostatic potential measured across DPPC. The application of an external electric field influences the transmembrane potential of a biological cell [75]. If the magnitude of the induced potential is significantly higher, it engenders poration at membrane surface. The time resolution of electroporation is in sub microsecond [76]. The analysis of electric potential across DPPC reflects the force or stress imbalance across the model simulation [77]. GROMACS utilizes the potential on the left side of the simulation box as its reference for evaluating the electrostatic potential. For this model, the potential was evaluated by aggregated charge distribution across distributed 10 slices of the simulation box. The integration of this charge distribution was used to derive the electrostatic potential. In general, the electric potential, V of a point charge, q can be calculated from the following equation (7):

$$V = \frac{k}{r} \int q \, dt \qquad (7)$$

where, r is the distance from the charge for which the potential is evaluated, and k is a constant $=9.0 \times 10^9 \ N.m^2/C^2$.

From Fig. 9 it is evident that the primary reference potential was 500 mV which was evaluated across the center of the DPPC layer. This value is similar to data obtained from previous literature [78]. Yu et al. (2023) measured the membrane DPPC bilayer at 323.15 K and DPPC monolayer at 321 K [87] . The potential was measured as a function of distance and the dipole potential measured from water phase to hydrophobic core was approximately 0.4 V. Experimentally the potential is measured by placing electrodes and values are approximately 0.3 - 0.4 V [87, 88]. The results are in close approximation to our initial electrostatic potential of DPPC layer of 0.5 V but on the onset of external applied field the electrostatic potential across the hydrophobic region escalates to a higher magnitude. This was due to the difference in ion concentration across the membrane. The application of an external field effects the internal polarization due to the dielectric nature of the cell membrane. This causes the molecules to rotate and align in the direction of the applied field (z-direction) and can be quantitatively measured from its electric dipole moment as shown in Fig. 10.

The electrostatic potential of DPPC is slightly less than 2V for an applied field of 0.4 V/nm. However, higher field strengths account for a greater change in transmembrane potential across DPPC allowing the hydrophobic region to create temporary defects. The results are in close approximation with literature where the measured transmembrane potential is approximately 3V to create a pore size of 10 nm [34]. The discrepancy between the results of Tieleman (2004) can be explained due to the size of the simulation model which has impact on the required threshold to induce nanopores. Application of an external field causes an average force, F acting on the simulation model and the magnitude can be calculated using equation (8):

$$F = Q\frac{E}{l_z} \tag{8}$$

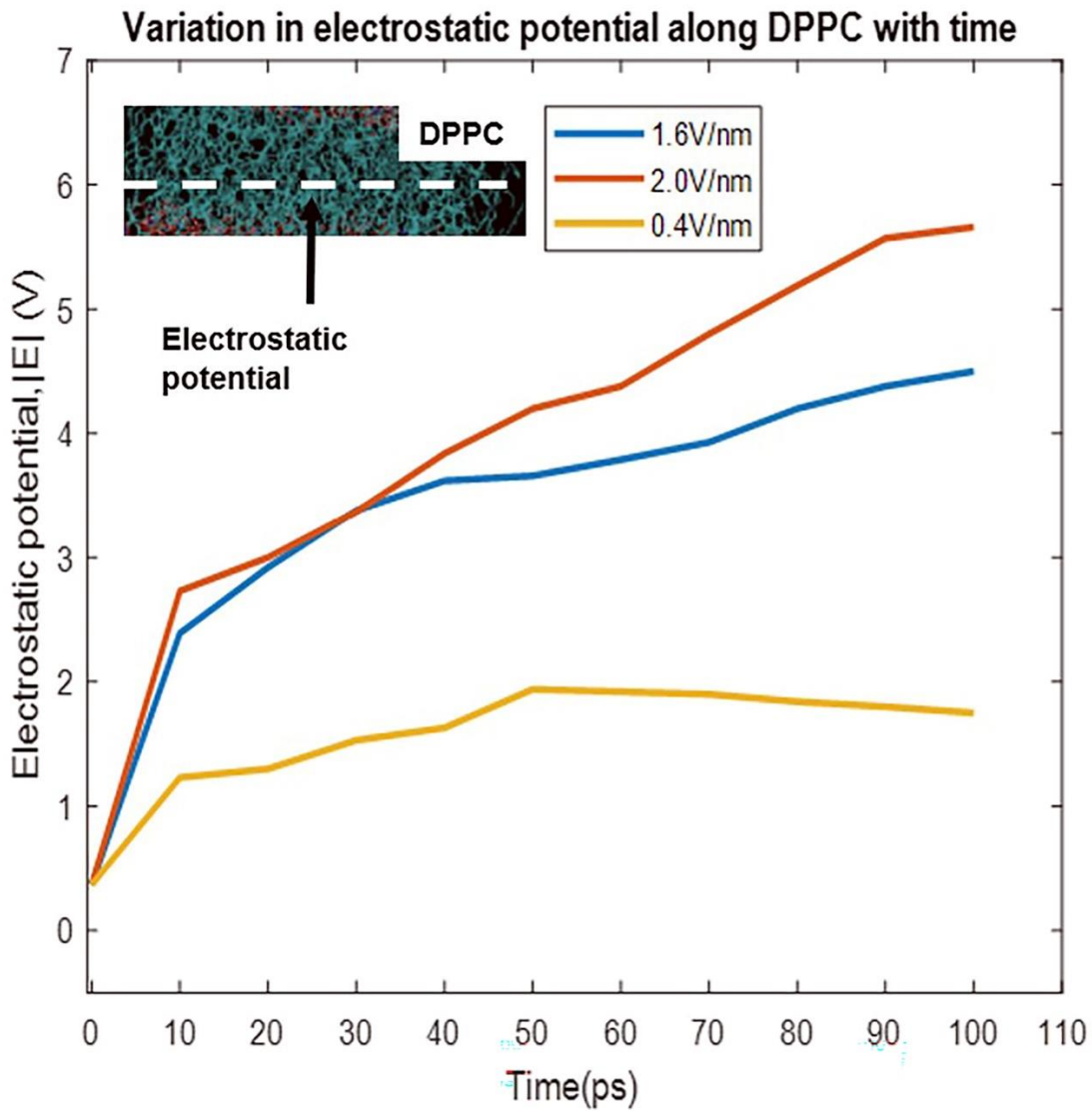

**Fig. 9: The change in the magnitude of electrostatic potential along the DPPC region. With the increase in magnitude of the field the potential simultaneously increases until saturation.**

where Q is the cumulative charge on the system, E is the magnitude of the applied field and $l_z$ is the length of the DPPC layer. From Fig. 9 for field magnitude of 0.4 V/nm is not equivalent or greater than in order to overcome the membrane impedance. Hence, membrane poration was not realized and this is evident that a higher electric field magnitude is obligated for both influencing

DPPC potential and water/ion penetration in the hydrophobic region. Increasing the applied field fourfold to 1.6 V/nm encourages nanopore formation which is evident from the graph in Fig. 8. It is also realized that the electrostatic potential across DPPC z-region increases which accounts for increased membrane potential. In literature, experiments revealed that when cell are exposed to an external field that causes their induced membrane potential to increase between a threshold magnitude it eventually causes the membrane permeability to increase [79-81]. For this particular simulation, by comparing the data from Fig. 8 and Fig .9 respectively which shows that water penetration in DPPC layer is not significant for 0.4 V/nm applied electric field and the corresponding induced membrane potential is not equivalent for overcoming the hydrophobic nature of DPPC layer. But on increasing the field strength to 1.6 V/nm, the corresponding force increases proportionally.

Fig. 10 shows the field strength is proportional to the dipole when measured along the DPPC region. This increases the charge separation between the DPPC molecules, leading to formation of nanopores. The electrostatic potential also increases, and it reaches its threshold range at 30 ps because water penetration therefore accelerates. Pavlin and Miklavcic (2003) used the solution of Laplace equation to derive equation for equivalent conductivity of permeabilized cell. However, they used a dipole approximation for their analysis. Hence a concomitant computer simulation to obtain the dipole magnitude can be effective for equivalent conductivity analysis.

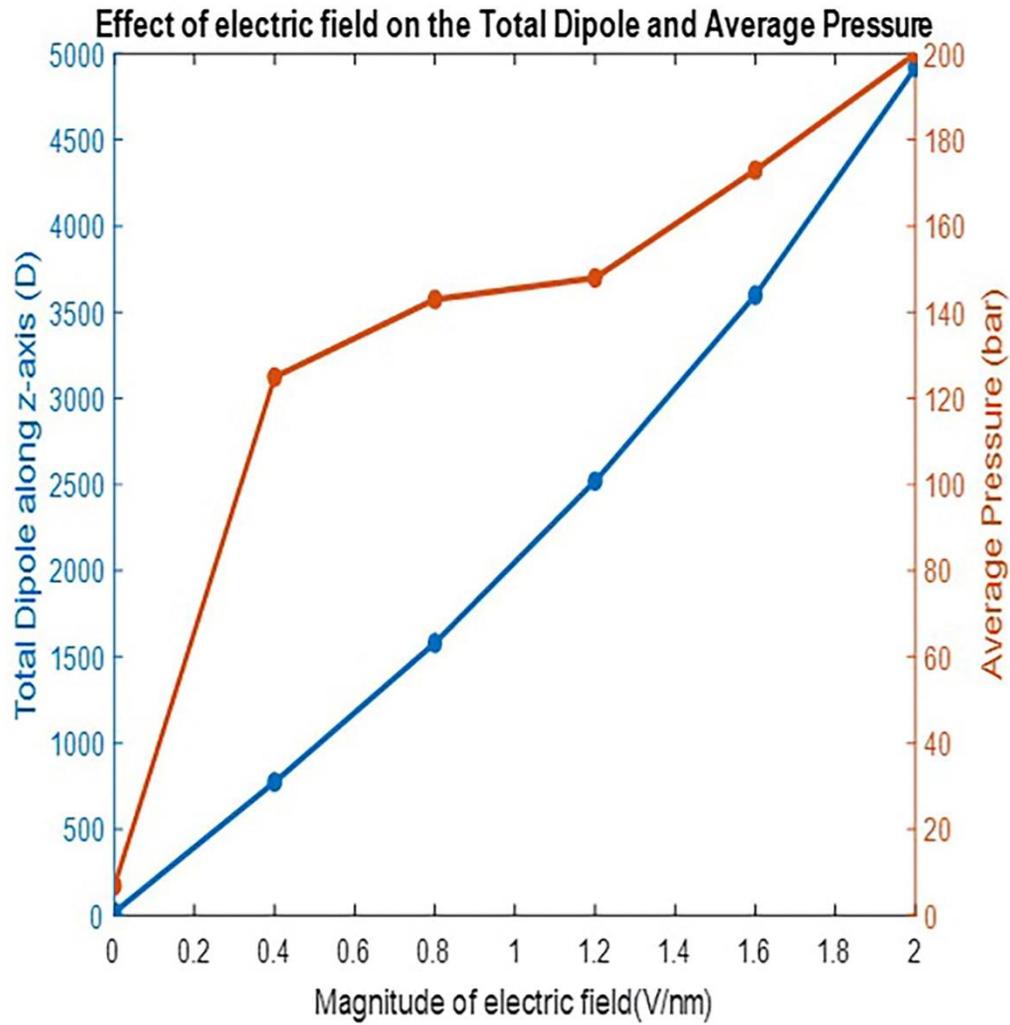

**Fig. 10:** The relation between dipole (D), Pressure (bar) and applied Electric field (V/nm). The relation between electric field and polarity (blue plot) appears linear from the simulation data. The average pressure shows an increasing pattern with the external electric field applied (red).

**Conclusion and Future Applications**

This paper highlights on the analysis of transmembrane potential change and conductivity of an isolated cell under external electric field perturbation. *Pauly and Schwan (1959), Schwan (1983) and Grosse and Schwan (1992)* proposed models on evaluating the transmembrane potential change are all discussed with their limitations. The membrane conductance is an important parameter for electroporation that leads to enhanced membrane permeability. The induced membrane potential drop is correlated to external conductivity, and if the external conductivity decreases a higher magnitude of electric pulse is required to induce the same membrane permeability.

Secondly, a MD simulation is performed to obtain a time dependent function change in permeability for a varied range of electric field. The critical characteristic threshold is determined for the model and an attempt is made to correlate with two parameters: (i) *change in electrostatic potential of DPPC (ii) dipole of the system*. The primary "resting" reference potential for our simulated model was 500 mV. Upon application of the external field, the electrostatic potential of DPPC simultaneously increased until a critical value of induced potential is obtained after which it attains saturation. This is explained because a further escalation of electro induced potential difference is limited by a decrease in the membrane permeability factor. Hence, the membrane permeability factor is an important parameter that forms the *"boundary"* between reversible and irreversible electroporation. *Pavlin and Miklavcic (2003)* emphasized on the dipole term in their analytical calculation of effective conductivity using the equivalent principle. Hence, the MD simulation allows to obtain the magnitude of dipole term for various magnitude of electric field. This can account for a more accurate analytical calculation of the effective conductivity and induced transmembrane potential.

Magnetoelectric smart composites such as cobalt ferrite-barium titanate (CFO-BTO) has shown potential to influence the membrane permeability due to nanoparticle-cell interaction [82-86]. The magnetoelectric materials can be externally controlled by alternating magnetic field and when brought in proximity of the cell will generate corresponding electric pulses that can influence the conductivity and transmembrane potential of the cell as discussed earlier in this paper. For future work, an understanding of the generated electric field magnitude for magnetoelectric materials can be studied and how they impact the transmembrane potential of the cell to enhance the cell's permeability for drug delivery.

**Acknowledgement**

The author(s) acknowledge the discussion with Dr. Qin Hu, Associate Professor at Eastern Michigan University for her guidance on MD simulation using GROMACS. The author also acknowledges the contribution of Dr. Ahmed Abdelgawad, Professor at Central Michigan University in the preparation of this manuscript.

The author also acknowledges the scholarship grant from School of Engineering and Technology, Central Michigan University, during the study.

**Conflict of Interest**

The author(s) have no conflict of interest.